\documentclass[prb,twocolumn,amsmath,amssymb,showpacs,letterpaper]{revtex4-1}
\usepackage{graphicx}
\usepackage{epstopdf}
\usepackage{epsfig}
\begin{document}
\title{A real-time software simulator for scanning force microscopy}

\author{V. Chandrasekhar}
\email{v-chandrasekhar@northwestern.edu}
\author{M. M. Mehta}
\affiliation{Department of Physics and Astronomy, Northwestern University, Evanston, Illinois, 60208, USA}



\begin{abstract}
We describe software that simulates the hardware of a scanning force microscope.  The essential feature of the software is its real-time response, which is critical for mimicking the behavior of real scanning probe hardware.  The simulator runs on an open-source real time Linux kernel, and can be used to test scanning probe microscope control software as well as theoretical models of different types of scanning probe microscopes. We describe the implementation of a tuning-fork based atomic force microscope and a dc electrostatic force microscope, and present representative images obtained from these models.  
\end{abstract}

\maketitle
Obtaining and interpreting images with a scanning probe microscope is a complicated task, even more so in the case of completely home-built microscopes with both custom electronics and custom software.  This is because of the integral relationship between software and hardware in scanning probe microscopes, particularly in the case of more recent instruments, where many of the functions previously performed by analog electronics are now performed in software, either on the control computer itself or on a separate digital signal processor.  During development, in the frequent case that something goes wrong, it is not easy to identify whether the problem lies with the software or the hardware.  Thus, it would be useful to have a diagnostic tool to test whether the control software is indeed performing as designed.

The heart of such a software simulator is in principle very simple:  one needs only to convert the position of the tip (as output by the scanning probe control program, perhaps by means of a voltage) to a voltage corresponding to a feedback signal that can be read back by the control program.  This conversion occurs according to some model of the tip-sample interaction, which in turn depends on the type of scanning probe microscopy, be it atomic force microscopy (AFM), scanning tunneling microscopy (STM), magnetic force microscopy (MFM) or electrostatic force microscopy (EFM), to name a few.  The difficult part is to do this in a time-critical manner; the response of the simulator must be as fast or faster than typical scanning probe hardware, which means the response must be in fractions of a millisecond, and the response must be deterministic, in that response loops must be executed when expected.  This is usually beyond the capabilities of standard desktop computers running modern operating systems, as 
these operating systems have preemptive multitasking, making their response times at best on the order of 10 ms, and more critically, not deterministic.  The software we describe here\cite{github} is based on an open-source, real-time version of Linux running on a standard desktop computer that enables loop times as short as 25 $\mu$s (more reliably around 50 $\mu$s), corresponding to an upper limit frequency response on the order of 20 kHz, fast enough to model most commercial and home-built scanning probe electronics.  
Another major advantage of such a software simulator is that one is able to easily test different models of tip-sample interactions as well as the influence of various model parameters on the resulting images.

The remainder of the paper is organized as follows:  Section I gives an overview of the software and hardware development platform.  Section II describes the model used to implement an atomic force microscope (AFM), and Section III describes the model used to implement an electrostatic force microscope.  These are two modes that are important for our own research, and we emphasize that many other types of interactions can also be implemented relatively easily if appropriate theoretical models are available.

\section{Software and hardware description}

The development environment that we have used for the SPM simulator program is similar that used to develop our real-time scanning program RTSPM.\cite{chandra}  The details can be found in Ref. \citenum{chandra}, so we shall only give an outline here.  We emphasize that all the software used is open-source.  In order to achieve real-time control, we use a desktop computer running a Linux kernel patched with the Real Time Application Interface (RTAI).\cite{rtai}   Communication with the data acquisition hardware is achieved through the open-source Comedi drivers\cite{comedi} with appropriate real-time extensions that interface well with RTAI.  For this paper, we used a National Instruments\cite{ni} PCI-6052E card for input and output, although any of the data cards supported by Comedi can be used, so long as they meet the required data acquisition rates of the program.  The PCI-6052E card has analog input and output rates of 333 kS/sec, more than sufficient for our purposes, but we have also used other 
National Instruments data acquisition cards.

\begin{figure*}[!]
\includegraphics[width=16.0cm]{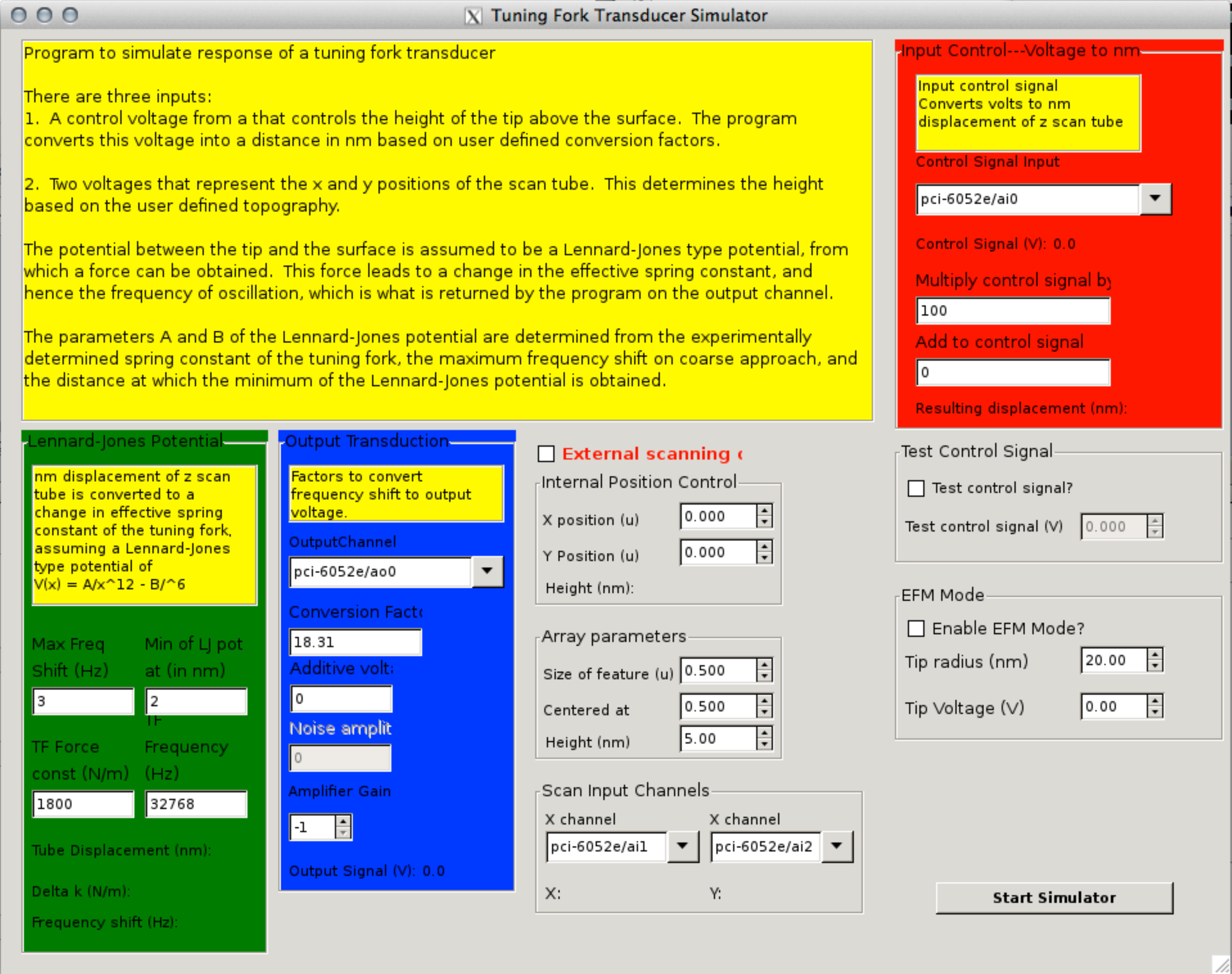}
\caption{Screenshot of the front panel of the program.}
\label{fig1}
\end{figure*}

The real time components of the program are coded in a library as callable subroutines that are written in C using the integrated development environment Code::Blocks,\cite{codeblocks} which is also open-source.  The graphical user interface (GUI) is written in Free Pascal\cite{freepascal} using the open-source integrated development environment Lazarus;\cite{lazarus} the real-time parts of the program are called by this main program.  
Figure \ref{fig1} shows a screenshot of the GUI of the program, which lets the user choose the channels for the data acquisition inputs and outputs.  There are three data acquisition inputs to the program, corresponding to the $x$, $y$ and $z$ voltage drives of a standard piezotube scanner, which can be selected independently from the GUI.   These inputs are provided by a SPM control program:  in our case, this control program is the real-time control program RTSPM that we developed earlier.\cite{chandra} The voltages on the $x$ and $y$ channels are assumed to correspond directly to the $x$ and $y$ displacements of the scanner:  of course, one can easily implement an appopriate scale factor for each channel independently to model specific hardware if needed.  For the $z$ voltage, the user can input the scale factor directly.  For the data shown in this paper, we use a scale factor of 155 nm/V, corresponding to the scan tube in our physical instrument.  For testing purposes, reading of the $x$ and $y$ 
channels can be disabled, and the $x$ and $y$ positions of the scanner can be manually entered from the GUI.    

As we noted earlier, the real-time part of the program is very simple in concept.  Once the simulator is started, a real-time loop runs continuously with a loop time of 50 $\mu$s.  In the loop itself, the program first determines the $x$ and $y$ positions of the scanner.  If external scanning is enabled, these positions are read directly from the $x$ and $y$ input channels, otherwise they are taken to be the manually entered values discussed above.  Depending on the $x$ and $y$ positions of the scanner, the program then determines what the corresponding topographical height should be.  For this paper, we have taken the simplest structure, an array of squares of specified lateral dimension and specified height with a lattice constant of 1 $\mu$m, although clearly more complicated topographic structures can easily be programmed.   The program then reads the $z$ channel to determine the height of the scanner.  Since the program now knows the height of the $z$ scanner and the topographic height of any feature at 
that $x-y$ position directly below it, it can then calculate and output a voltage according to whatever model is being used to generate the tip-sample interaction.  This output voltage, which corresponds to the feedback signal input to the SPM controller program, is then read by the SPM controller program to adjust the $z$ piezo voltage accordingly.  The process is then repeated on the next 50 $\mu$s cycle.

The voltage that the program generates depends on the model of the tip-sample interaction.  We discuss below two models that we have implemented, although other tip-sample interactions can also be modeled.

\section{Atomic force microscope}
Our own research is devoted to development of a low temperature scanning probe microscope, so it is natural for us to first try to implement a model for an atomic force microscope.  Our home-built scanning probe microscope is based on a tuning fork transducer,\cite{seo} so the parameters in the model described below will refer to the measured parameters from this instrument, but the program allows these parameters to be modified to match any force transducer.  
 
We start by assuming that the interaction between the tip and sample is due to van der Waals' forces at larger distances with a strong repulsion at short distances.  This interaction can be described by a Lennard-Jones type potential of the form
\begin{equation}
V(z) = \frac{A}{z^{12}} - \frac{B}{z^6},
\label{eqn1}
\end{equation}
where $z$ is the distance between the tip and the sample, and $A$ and $B$ are constants.  The first term describes the strong repulsion between tip and sample at very short distances, and the second describes the relatively short range van der Waals' interaction.   The two unknowns in the potential are the parameters $A$ and $B$.  We shall use the measured characteristics of the close-approach curve to determine these constants.   

The force between the tip and the sample corresponding to this potential is 
\begin{equation}
F(z) = - \frac{\partial V(z)}{\partial z} = \frac{12 A}{z^{13}} - \frac{6 B}{z^7}.
\label{eqn2}
\end{equation}
To eliminate one of these constants, we specify the position of the minimum of force $F$ as a function of $z$ as $z_0$.  By setting $\partial F(z)/\partial z = 0$ at $z=z_0$, this allows us to express $B$ in terms of $A$ and $z_0$,
\begin{equation}
B = \frac{2 A}{z_0^6}
\label{eqn3}.
\end{equation}
A tuning fork based AFM is operated in non-contact mode.  The tuning fork is oscillated at its resonance frequency $f_0$, which for our tuning forks is close to 32768 Hz.  The tip-sample interaction modifies the resonant frequency.  The shift in resonant frequency $\Delta f$ can be thought of as arising from a shift $\Delta k(z)$ in the effective spring constant of the tuning fork
\begin{equation}
\frac{\Delta f}{f_0} = \frac{1}{2} \frac{\Delta k (z)}{k_0},
\label{eqn4}
\end{equation} 
where $k_0$ is the spring constant of the tuning fork far from the surface.  This is because the resonance frequency is proportional to the square root of the spring constant.  We shall use $k_0$=1800 N/m, corresponding to the spring constants of the tuning forks we normally use.

Now $\Delta k(z)$ is given by the derivative of the tip-sample force
\begin{equation}
\Delta k(z) = - \frac{\partial F(z)}{\partial z} = 12 A \left[ \frac{13}{z^{14}} - \frac{7}{z_0^6 z^8} \right].
\label{eqn5}
\end{equation}

In non-contact mode, one can measure the amplitude or phase of the oscillation as a feedback signal.  However, if the quality factor $Q$ of the force transducer is large, as it is for the tuning forks, it may take a very long time for these signals to relax to their proper values.  Hence, it is common to use a phase-locked-loop (PLL) to stay on the resonance and track the change in resonant frequency as a function of distance $z$.  Thus, we will use the change in frequency as the feedback signal for our SPM controller.

The change in frequency is proportional to $\Delta k(z)$, which still has one unknown constant $A$ (Eqn. \ref{eqn5}).  Since one directly measures the frequency-distance curve during close approach, $A$ can be determined from this curve.  Let the total change in frequency between where the tip is far from the sample ($z\rightarrow \infty$) and the value of $z=z_{min}$ where the frequency shift $\Delta f$ has its minimum be $\Delta f_0$.  ($z_{min}$ is related to $z_0$ by $z_{min} = 1.217 z_0$.)  Then $A$ can be expressed as
\begin{equation}
A = 0.2674  \Delta f_0 \frac{k_0}{f_0} z_0^{14} = 0.0171   \Delta f_0 \frac{k_0}{f_0} z_{min}^{14}.
\label{eqn6}
\end{equation}       
Determining $z=0$ in a real experiment (and hence the absolute values of $z_0$ or $z_{min}$) is difficult, since $z=0$ is the point at which the tip makes contact with the sample, and one prefers not to crash the tip into the sample.  At distances far from the surface ($z\rightarrow \infty$), $\Delta f$ is ideally 0:  in reality, due to experimental offsets, it may have a small finite value, which we denote $\Delta f_\infty$.  Then the value of $z$ near the surface at which $\Delta f = \Delta f_\infty$  is by definition $z_0$, and the minimum of 
$\Delta f$ as a function of $z$ occurs at $z = z_{min} = 1.217 z_0$.  Thus
by measuring the value of $z$ at these two points, one can
determine $z_0$.
Consequently, by specifying the known or measured quantities $\Delta f_0$, $k_0$, $f_0$ and $z_0$, one can determine all the parameters of the model.  These quantities are entered in the main program by hand.  The frequency shift, and hence the feedback signal that the program generates, can then be calculated using Eqns. (\ref{eqn4}), (\ref{eqn5}) and (\ref{eqn6}).

\begin{figure}[!]
\includegraphics[width=8.5cm]{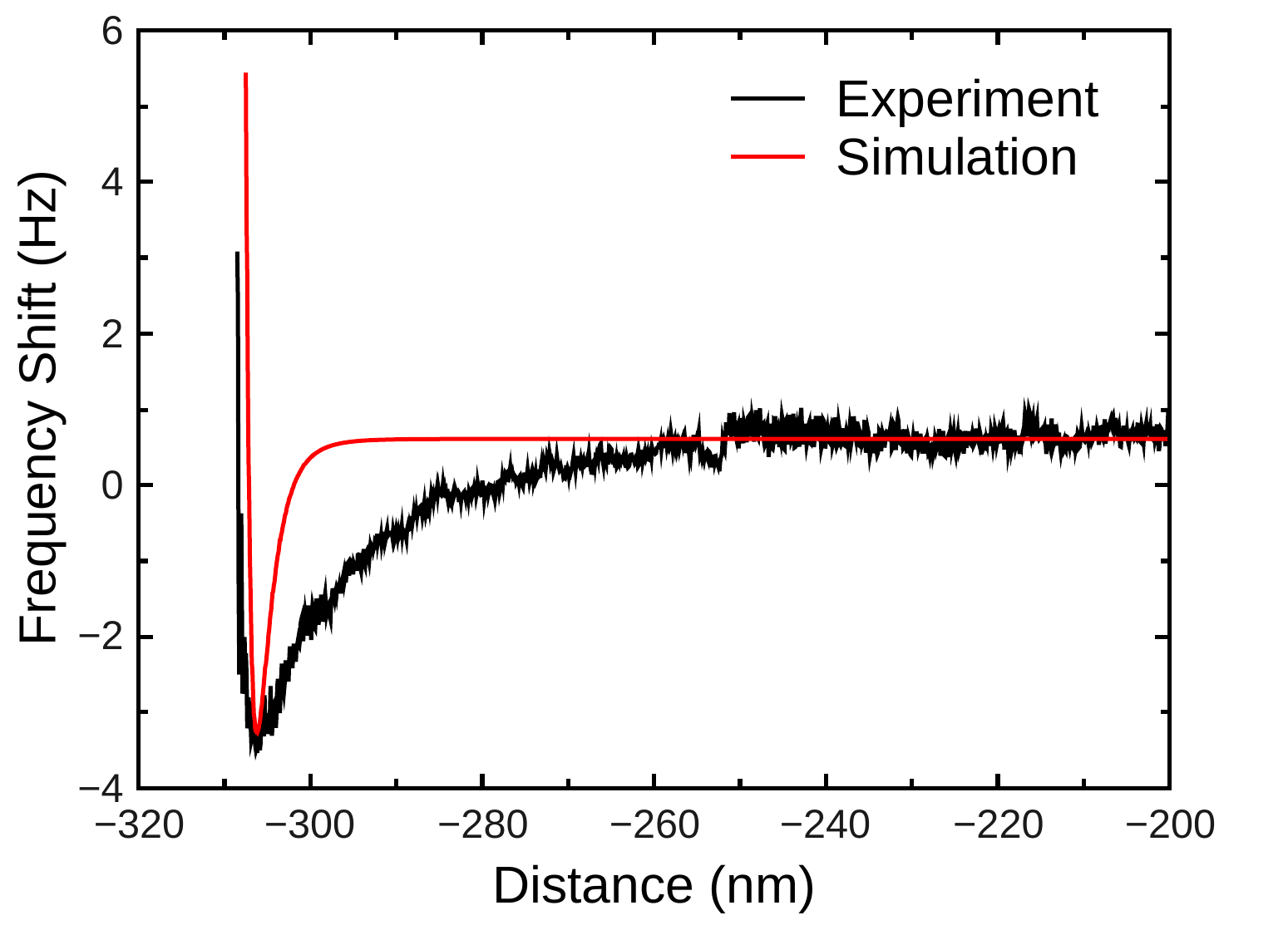}
\caption{Black: Experimental close-approach curve obtained with a tuning fork transducer with an attached 50 $\mu$m etched tungsten wire acting as a tip.  The spring constant of the tuning fork is 1800 N/m, and the curve was obtained with the RTSPM program.  Red:  Approach curve obtained with the simulator program, using a Lennard-Jones type potential as described in the text. }
\label{fig2}
\end{figure}

The black trace in Fig. \ref{fig2} shows an example of an experimental close-approach curve measured using our home-built  tuning fork scanning probe microscope and the SPM control program RTSPM.  From this curve, we determine $\Delta f_0 = 4.06$ Hz and $z_0=9.22$ nm.  The red trace in Fig. \ref{fig2}(b) shows the corresponding curve generated using the SPM simulator program, again taken with the SPM control program RTSPM.  It can be seen that the experimental close-approach curve is broader than the simulation based on the Lennard-Jones potential.  We do not know the origin of this discrepancy; however, it should be noted that the experimental curve was taken with a conducting tip, and hence may be influenced by residual electrostatic interactions, which have a slower power-law dependence.     
\begin{figure}[!]
\includegraphics[width=8.5cm]{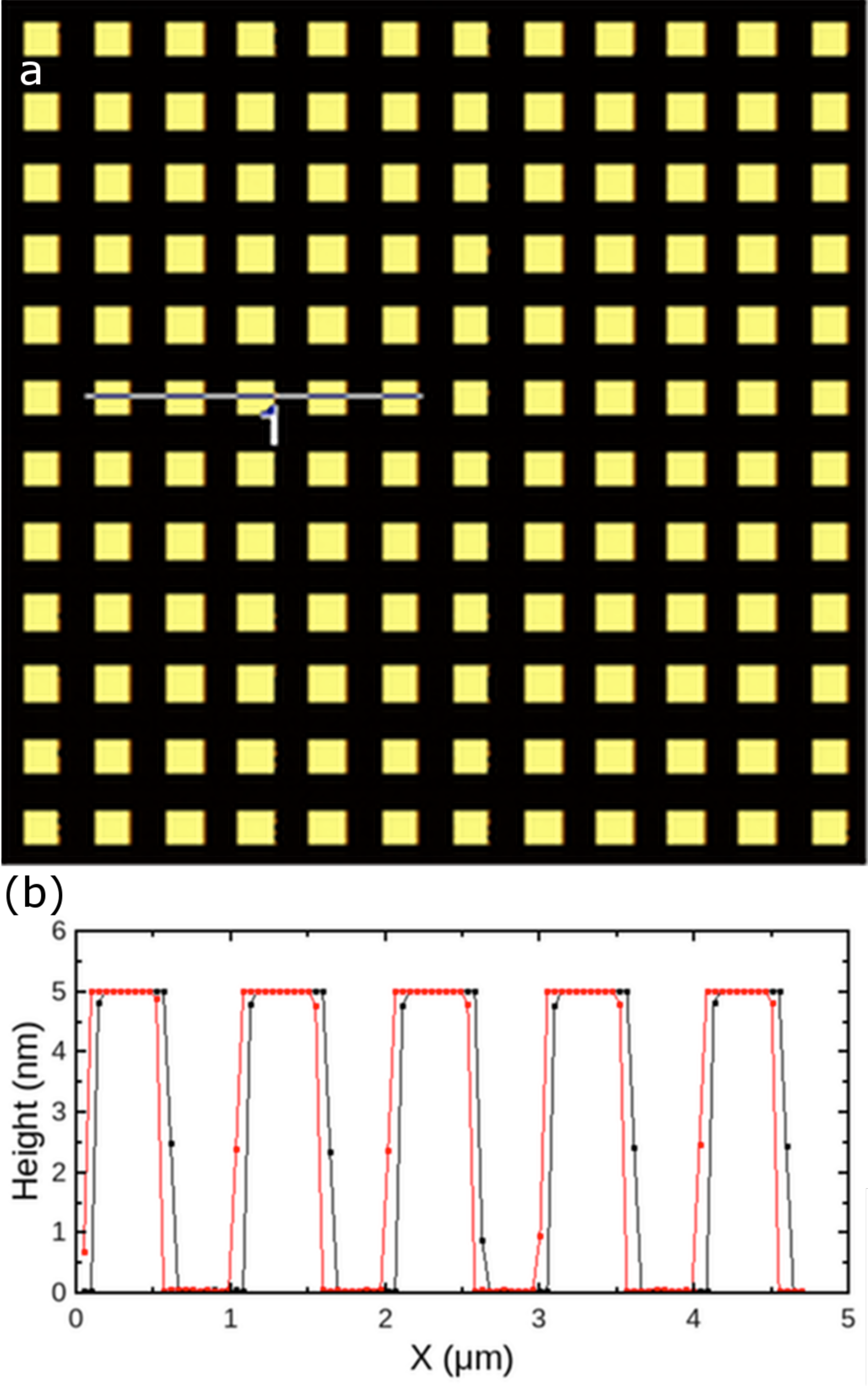}
\caption{(a)  12 $\mu$m $\times$ 12 $\mu$m forward topographic scan with a resolution of 256 x 156 pixels.  Each square in the image is 0.5 $\mu$ $\times$ 0.5 $\mu$m and has a height of 5 nm.  (b) Line scan profiles corresponding to the line marked in (a) for the forward scan (black trace) and the reverse scan (red trace).  }
\label{fig3}
\end{figure}
Figure \ref{fig3}(a) shows an image generated by the RTSPM program coupled to the SPM simulator program.  (This figure and other images in this paper are generated using the open source SPM analysis program Gwyddion.\cite{gwyddion})  As described earlier, the ``sample'' is a computer generated array of squares of side 0.5 $\mu$m and height 5 nm, separated by a distance of 1 $\mu$m.  The RTSPM has a real-time proportional-integral-differential (PID) controller\cite{chandra} that controls the extension of the scan tube based on the desired set point, which is a fixed frequency deviation from the $z\rightarrow\infty$ limit.  For the image in Fig. \ref{fig3}(a), the PID parameters used in the RTSPM program were $P=5 \times 10^{-5}$, $I=0.08$ ms, and $D=0.001$ ms with a set point of 2 Hz, which puts us in the hard repulsive region of the approach curve.  In this region, a small change in $z$ gives rise to a large change in frequency.  Nevertheless, as can be seen from the line profiles shown in Fig. \ref{fig3}(b),
 the topography is accurately mapped by the RTSPM program in both the forward and reverse scan directions, with a difference corresponding to about 1 pixel.  The scan resolution is 256x256 pixels, and the RTSPM program averages for 10 ms on each pixel to reduce the noise, so the image of Fig. \ref{fig3}(a) took approximately 12 minutes to acquire.   

\begin{figure}[!]
\includegraphics[width=8.5cm]{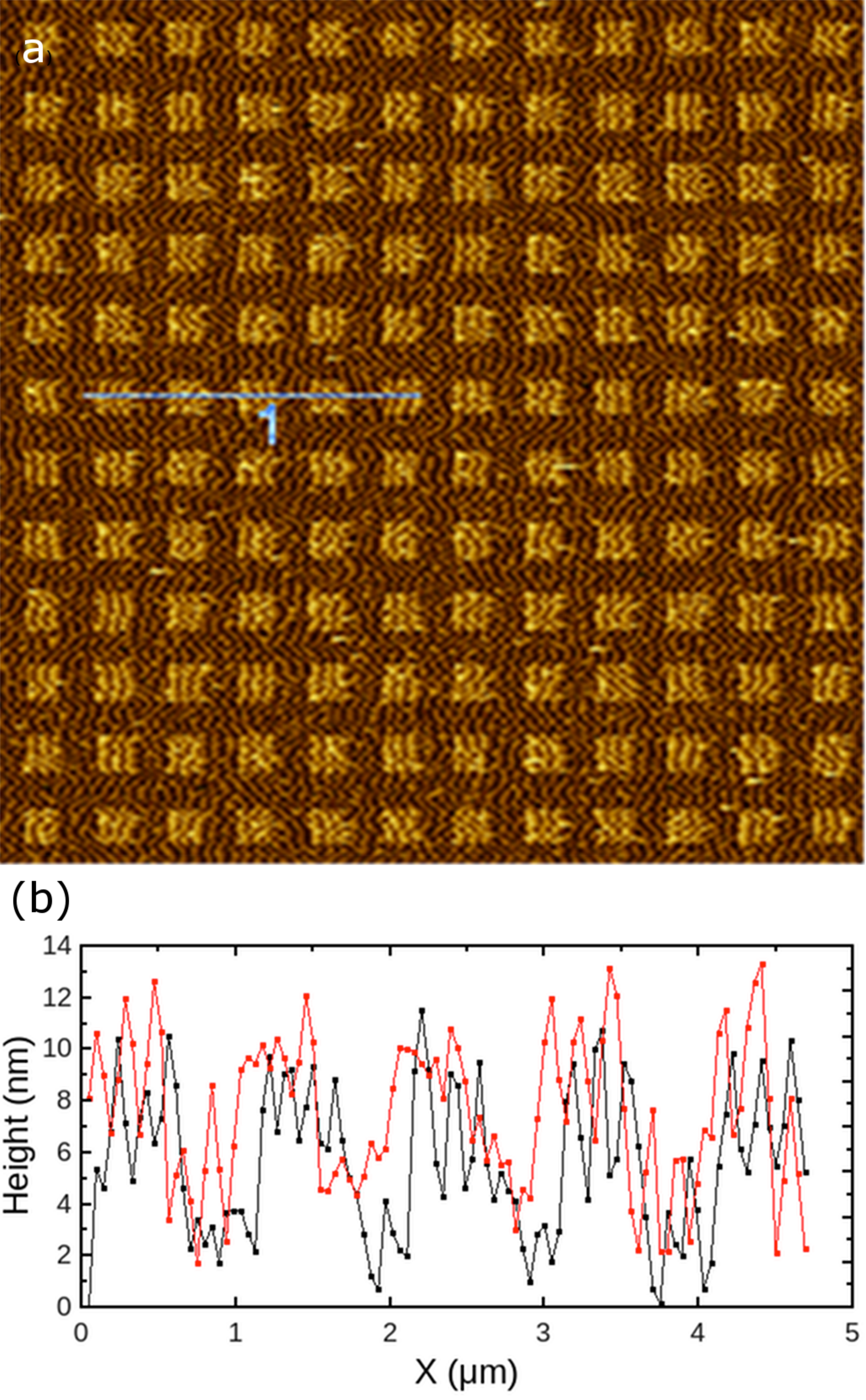}
\caption{(a)  Same scan as in Fig. \ref{fig3}, but with a non-realtime simulator coded in LabView on a computer running Windows XP.  (b) Line scan profiles corresponding to the line marked in (a) for the forward scan (black trace) and the reverse scan (red trace).  }
\label{fig4}
\end{figure}
In order to illustrate the importance of having a real-time SPM simulator program, we have also encoded the same program on a desktop computer running Windows XP in National Instruments LabView without any real-time extensions.  In this case, even on a relatively fast computer, the minimum loop time is at best 10 ms.  Figure \ref{fig4}(a) shows the image obtained with the RTSPM program using the same PID parameters and setpoint as in Fig. \ref{fig3}, but coupled to the LabView simulator.  It can be seen that the image quality is much worse, and the line profiles in Fig, \ref{fig4}(b) show that this is because the $z$ position of the scanner does not follow the topography.  Since all parameters in the RTSPM program are the same for Figs. \ref{fig3} and \ref{fig4}, the difference is clearly due to the lack of real-time response in the LabView program.  

\section{Electrostatic force microscope}

The van der Waals' interaction for AFM in principle can be extended to other types of tip-sample interactions.  For example, for scanning tunneling microscopy, one can model a tunneling current that depends exponentially on the distance between the tip and the sample, and also on bias.  For a more sophisticated model, one can think of a conducting ``sample'' with a spatially inhomogenous density of electronic states.  Even relatively inexpensive modern multicore desktop computers should be able to calculate the feedback response of most tip-sample interaction models in a loop time of 50 $\mu$s.

For long range tip-sample interactions, the models are more complicated.  For example, for magnetic force microscopy (MFM) on a ferromagnetic film, one must consider not only the magnetic interaction between the magnetic tip and the part of the sample just below the tip, but also the magnetic interaction with other parts of the ferromagnetic film that are much further away.  In electrostatic force microscopy (EFM), the electrostatic interaction between the tip and the sample is also long range, and an accurate calculation would require numerical techniques.  There have been a number of papers that have developed approximations for the tip sample interaction both in the context of EFM and scanning capacitance microscopy.\cite{hudlet,saintjean,gomez,naitou}  However, with some simplifiying assumptions, one can use a simple model that gives a realistic reproduction of an electrostatic image.

The assumption that we shall make is that the conducting tip on the scan tube is very close to the conducting surface of the sample, so that the interaction between the tip and the sample can be approximated by the interaction between a small conducting sphere and an infinite conducting plane, where the radius $R$ of the small sphere is the same as the radius of the microscope tip.  As can be expected, this assumption neglects fringe fields at the edges of a conducting region in the sample, but the result should be accurate beyond a few multiples of the tip-sample distance from any edge.

Consider then the force between a conducting sphere of radius $R$ and an infinite grounded conducting plane.  This is a classic problem that can be solved by the method of images.\cite{smythe}  The force between the sphere and conductor is given by
\begin{equation}
F_C(z) = \frac{1}{2} V^2 \frac{\partial C}{\partial z},
\label{eqn7}
\end{equation}
where $V$ is the voltage difference between the sphere and plane (which in the case of real metals will also include any contact potentials), and $C$ is the capacitance between them.  Since $C$ decreases with increasing distance $z$, the force is always attractive.  Calculation of the force then reduces to calculation of the capacitance, which is given by the series\cite{smythe}
\begin{equation}
C= 4 \pi \epsilon_0 R \sinh(\alpha) \sum_{n=1}^{\infty} \frac{1}{\sinh(n \alpha)},
\label{eqn8}
\end{equation}
where $\cosh{\alpha}=1 + z/R$.  The resulting expression for the force is 
\begin{equation}
F_C(z) = 2 \pi \epsilon_0 V^2 \sum_{n=1}^{\infty} \frac{\coth(\alpha)- n \coth(n \alpha)}{\sinh(n\alpha)}.
\label{eqn9}
\end{equation}
Since we would like to calculate the derivative of $F_C(z)$ with respect to $z$ to calculate the shift in frequency along the lines of Eqns. \ref{eqn4} and \ref{eqn5}, it would be nice to have a simpler equation to work with.  Hudlet \textit{et al.}\cite{hudlet} have shown that the following expression closely approximates Eqn. \ref{eqn9}, with a maximum error of a few percent\cite{hudlet}
\begin{equation}
F_C(z) = -\pi \epsilon_0 \left[ \frac{R^2}{z(z+R)}\right] V^2.   
\label{eqn10}
\end{equation}
We shall use this equation in our calculations of the frequency shift $\Delta f_C$ due to the electrostatic interaction.  This frequency shift can be calculated from Eqn. \ref{eqn4} and the relation $\Delta k_C(z) = - \partial F_C(z)/\partial z$.  We obtain
\begin{equation}
\Delta f_C = -\frac{\pi \epsilon_0 R^2 V^2}{2} \frac{f_0}{k_0} \left[ \frac{2 z + R}{[z(z+R)]^2}\right].
\label{eqn11}
\end{equation}  
Unlike the equation for the van der Waals' interaction, we note that there are no free parameters in this equation. 

\begin{figure}[!]
\includegraphics[width=8.5cm]{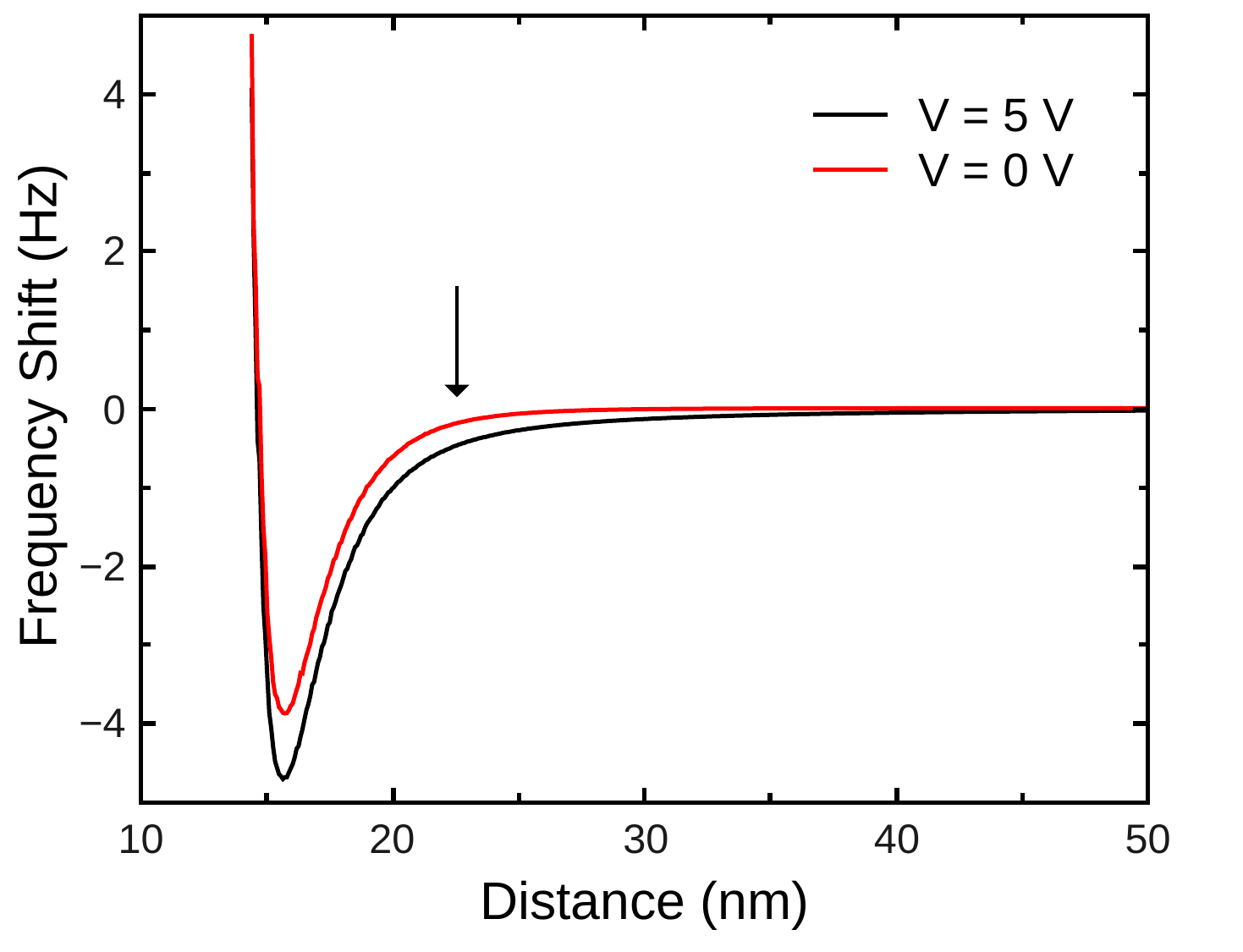}
\caption{Close-approach curves on a square with and without a voltage. The arrow indicates a distance of 8 nm from the point at which $\Delta$f = 2 Hz.}
\label{fig5}
\end{figure}

Figure \ref{fig5} shows the close-approach curves generated with the SPM simulator program both with and without the electrostatic force.  Here the radius $R$ of the tip is taken to be 20 nm, and the potential difference $V$ between the tip and sample is 5 V.  As is well known and can be seen from the two curves, the van der Waals' interaction is dominant close to the surface, at a distance of less than a few nm, while the electrostatic interaction, being longer range, contributes to the attractive potential at larger distances.  This forms the basis of two techniques to extract the electrostatic force.  In the first technique, for each line of the scan, a forward and reverse scan is made under feedback at a setpoint corresponding to a distance from the surface at which the van der Waals' force is dominant.  The resulting $z$ positions of the piezo scan tube reflect primarily the topography of the sample along the scan line.  The SPM controller then takes the scan tube out of feedback mode, raises the scan 
tube in the $z$ direction by a distance $h$, the lift height, and retraces the forward and reverse traces while recording the frequency shift.  Since the system is no longer under feedback, $h$ should be larger than any topographic features in the sample.  The idea behind this constant height technique is that for sufficiently large $h$, the primary contribution to the resulting signal will be from electrostatic forces.  One can also retrace the stored $z$ positions of the topographic trace at the added height $h$ while recording the frequency shift.  The hope is that this ``Lift Mode''  technique will subtract the contribution from the signal due to topography, leaving only the signal due to the electrostatic force.  Of course, one can also measure other signals such as the phase or amplitude of the oscillation when the system is not in feedback, but we have chosen here to measure the frequency shift as it is conceptually easier to understand.  The benefit of modeling the electrostatic interaction is that 
one can determine immediately what appropriate value of $h$ to use from the close-approach curves in Fig. \ref{fig5}.  If we use a set point of 2 Hz for the topographic image, it appears that a lift height of about 8 nm should give the largest electrostatic signal.   
\begin{figure}[!]
\includegraphics[width=8.5cm]{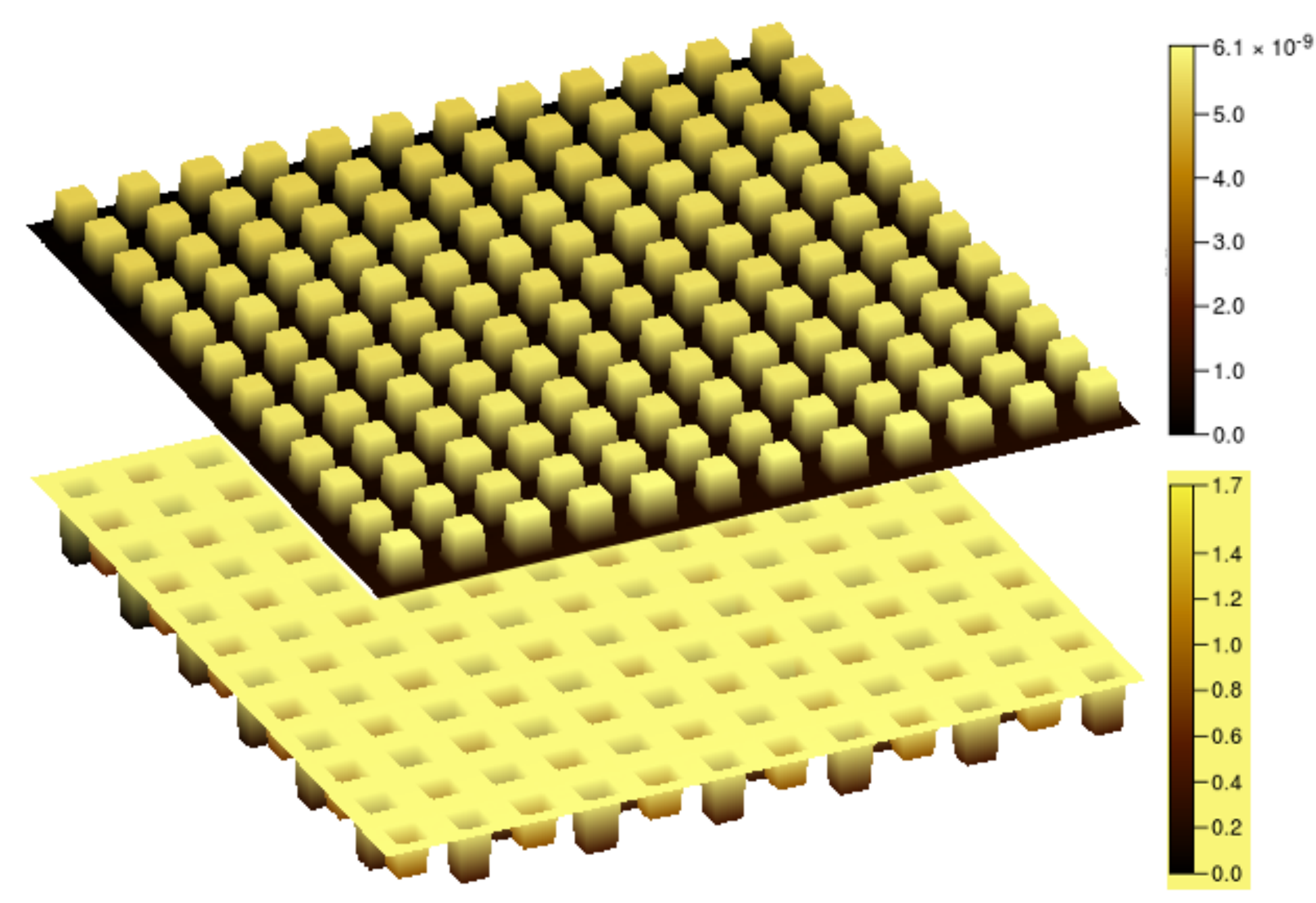}
\caption{Three dimensional topographic (top) and electrostatic force (bottom) images of the array of squares.  The EFM image was obtained using constant height mode, with a lift height of 8 nm.  The area of the scan is 12 $\mu$m x 12 $\mu$m, and the topographic height of each square is 5 nm. The colour bars are for the $z$-scale; for the topographic image the units are m, for the EFM images the units are Hz.}
\label{fig6}
\end{figure}
\begin{figure}[!]
\includegraphics[width=8.5cm]{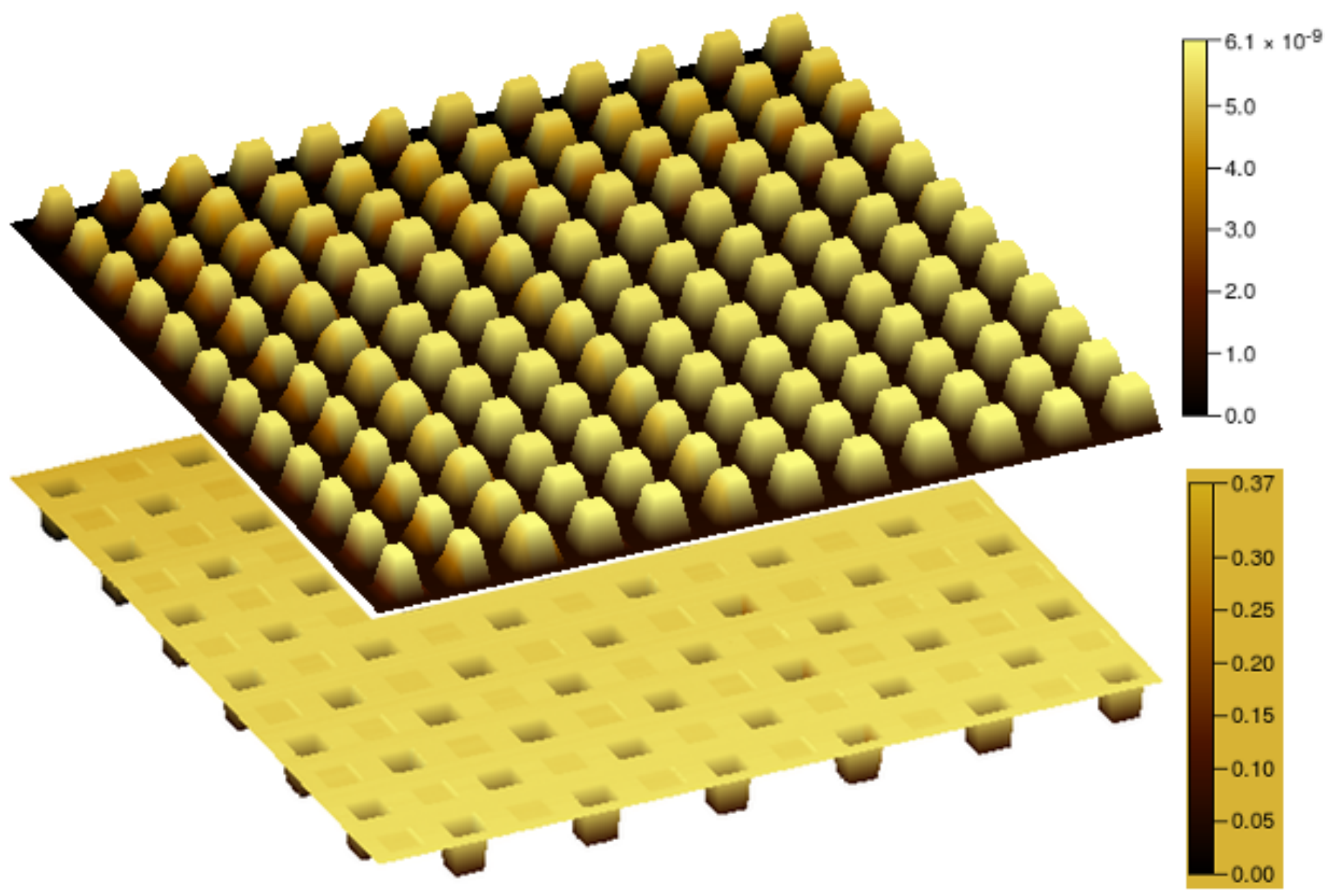}
\caption{Three dimensional topographic (top) and electrostatic force (bottom) images of the array of squares.  The EFM image was obtained using Lift Mode in the dc EFM module of RTSPM, with a lift height of 8 nm.  The area of the scan is 12 $\mu$m x 12 $\mu$m, and the topographic height of each square is 5 nm. The colour bars are for the $z$-scale; for the topographic image the units are m, for the EFM images the units are Hz.}
\label{fig7}
\end{figure}

In order to model an electrostatic sample, we use the same array of squares described earlier, but assume each alternate square in the array has a potential $V$ applied, with the tip and the other squares being grounded.  This approximates the standard samples frequently used for EFM calibration, which consist of interdigitated metallic fingers which have a potential applied only to every alternate finger, so that one can immediately distinguish between the topographic image and the electrostatic image.  Figure \ref{fig6} shows three dimensional representations of the resulting topographic and electrostatic images, using Constant Height mode in the dc EFM module of RTSPM.  For these images, we used a setpoint for the topographic image of 2 Hz (referring to Fig. \ref{fig5}), a voltage $V=5$ V, and a lift height of $h$= 8 nm.  
It is clear from the figure that there is substantial leakage of the topographic image into the EFM image.  This is not surprising, since for a lift height of 8 nm, the height of the tip above each square is only 3 nm so the van der Waals' force would contribute considerably to the signal. 

\begin{figure}[!]
\includegraphics[width=8.5cm]{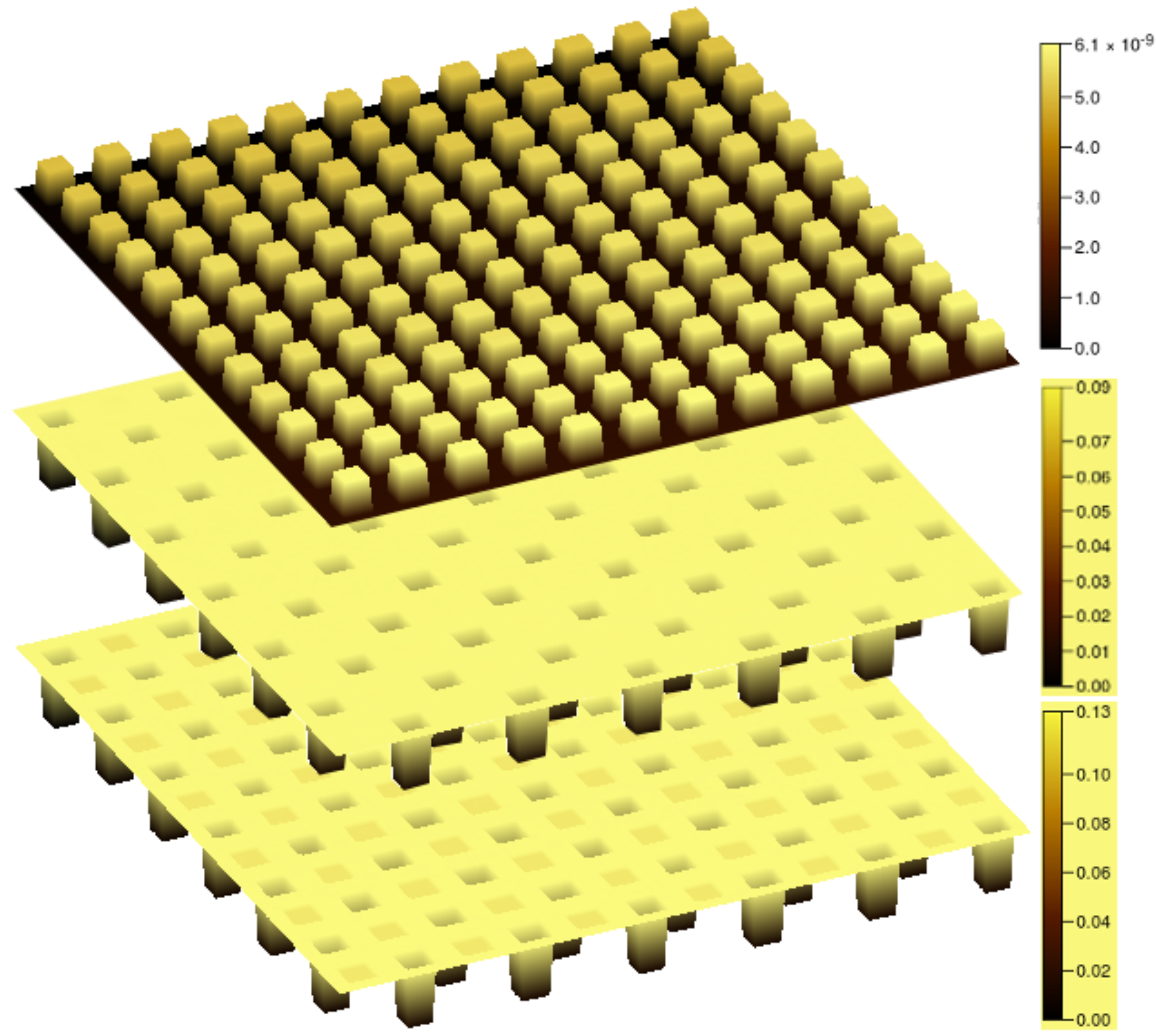}
\caption{Top: topographic image; middle: Lift Mode EFM image; and bottom: Constant Height EFM image, similar to Figs. \ref{fig6} and \ref{fig7}, but with a lift height of 20 nm. The Lift Mode and Constant Height Mode scan were acquired separately. The colour bars are for the $z$-scale; for the topographic image the units are m, for the EFM images the units are Hz.}
\label{fig8}
\end{figure}
For comparison, Fig. \ref{fig7} shows the corresponding image (at the same lift height of 8 nm) obtained using Lift Mode.  While there is some bleeding of the topographic image into the EFM image, it is quite small:  the ratio of the signal between squares with and without a potential is about a factor of 20.  The maximum frequency shift in Fig. \ref{fig7}(b) is about 300 mHz.  Clearly, Lift Mode is the preferable mode of operation when the sample has any appreciable topographical relief.  To eliminate any topographic signal in the EFM image, one can use a larger lift height.  Figure \ref{fig8} shows the Lift Mode and Constant Height mode EFM images for a lift height of 20 nm.  While the Constant Height mode image still shows some hint of the squares without any potential, the Lift Mode image shows no hint of the topography,  but only the electrostatic profile.  The overall Lift Mode signal is reduced in comparison to Fig. \ref{fig7}, but not by much (80 mHz peak frequency shift), a reflection of the long 
range nature of the electrostatic force. 

In summary, we have developed a real-time software simulator that models the response of a scanning probe microscope.  The simulator is useful for testing scanning probe control software as well as different models for tip sample interactions.     
  
\begin{acknowledgements}
This research was conducted with support from the US Department of Energy, Basic Energy Sciences, under grant number DE-FG02-06ER46346. 
\end{acknowledgements}

\end{document}